\documentclass[twocolumn,english,prl,showpacs,notitlepag]{revtex4}

\usepackage{graphicx}

\usepackage{amsfonts}

\begin{document}

\title{Coherent excitation transferring via dark state in light-harvesting
process}

\author{H. Dong, D. Z. Xu and C. P. Sun}

\email{suncp@itp.ac.cn}

\homepage{http://power.itp.ac.cn/ suncp/index-c.htm}

\affiliation{Institute of Theoretical Physics, Chinese Academy of Sciences, Beijing,
100190, China}
\begin{abstract}
We study the light absorption and energy transferring in a donor-acceptor
system with a bionic structure. In the optimal case with uniform couplings,
it is found that the quantum dynamics of this seemingly complicated
system is reduced as a three-level system of $\Lambda$-type. With
this observation, we show that the dark state based electromagnetically-induced
transparency (EIT) effect could enhance the energy transfer efficiency,
through a quantum interference effect suppressing the excited population
of the donors. We estimate the optimal parameters of the system to
achieve the maximum output power. The splitting behavior of maximum
power may be used to explain the phenomenon that the photosynthesis
systems mainly absorb two colors of light. 
\end{abstract}


\maketitle
To tackle the global problem of energy source \cite{energypro} people
may learn lots from the natural process of light-harvesting in plants,
algae and bacteria. The energy transfer mechanism of high efficiency
in light-harvesting process would help to design the new generation
of clean solar energy sources. Recently, the long-time coherent properties
of excitation in light-harvesting systems have been observed in experiments.
This coherence can be preserved in these structures for a long time,
even in room temperature. It seems that the high efficiency of light
conversion is related to these coherent properties even with quantum
natures. Therefore, the physical mechanism of photosynthesis assisted
by (somehow quantum) coherence attracts much more attention from both
experimental \cite{qcoherenceexp} and theoretical aspects \cite{YZhao,prb782008,venturi08cpc,Guzik09JPCB,mukamel09JCP,Thorwart09CPL,plenio09JCP,Guzik2010APL,Sarovar10NP,Ishizaki10NJP,CaoJPCB2010,Aspuru-Guzik_JCP08_09,SYangJCP2010}.

An inherent mechanism for high efficiency energy transferring may
be due to the optimizing of spatial structure of light-harvesting
systems. The X-ray analysis \cite{x-rays} has revealed some common
elements shared by different light-harvesting complexes in nature,
one of which is a ring structure with a centered reaction center \cite{HuBJ1998,Hu97PT}.
It is well-known that the nature selection rules always keep the most
adaptable feature in the biological system for the present environment.
Therefore, it is believed that this kind of structure takes advantage
in the light-harvesting process. The mechanism behind this optimal
structure may account for the high efficiency of the natural light-harvesting
process. Thus the investigation of mechanism of similar system would
be heuristic to design artificial light-harvesting systems with self-assembling
molecular array or the quantum dot array in top-down-semiconductor
fabrication in the future.

In this letter, we will study a generic model, which is similar to
the light harvesting complex of type I (LHC I), a centralized acceptor
surrounded by the coupled donors arranged in a ring. In natural light-harvesting
process, the light-capture process and excitation transferring happen
simultaneously. To mimic the natural process, we include light capture
process in the model by coupling the donors with photons in single
mode. By showing the present model with homogeneous coupling could
be reduced into the well-studied three-level $\Lambda$ system, we
find that the overall transfer efficiency is insensitive to the decay
of the donors when the eigen-frequency of the acceptor is resonant
with the light. This discovery implies that the dark state effect
suppresses the excitation population on the noisy donors so that the
energy transferring efficiency is dramatically improved. Otherwise,
the loss of donor excitations will largely decrease the transferring
efficiency to the acceptor.

\begin{figure}
 \includegraphics[width=7cm]{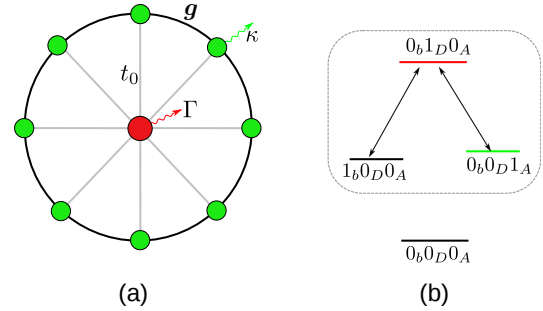} \caption{(\textit{Color online}) (a) The photon assisted donor-acceptor system
with dissipations in both donors (rate $\kappa$) and acceptor (rate
$\Gamma$); (b) The reduced energy spectrum in single excitation subspace
for its structure optimization}

\label{Fig1} 
\end{figure}

As illustrated in Fig.\ref{Fig1}(a), our model concerning light capture
and coherent excitation transfer, is similar to the structure of LHC
I. The one-dimensional circle array consists $N$ donors which is
also analogous to the ring structures in the LHC II, such as B800
or B850 ring. Similar to the pigment molecules in natural LHC I ring,
each donor can be modeled as a two-level system (TLS) $\left\vert e_{i}\right\rangle $
and $\left\vert g_{i}\right\rangle $ with energy level spacing $\epsilon_{i}\left(i=1,...,N\right)$.
TLS is a proper approximation for single excitation case in practice.
In most natural conditions, the LHCs always have only one excitation
\cite{Sarovar10NP}, thus our discussion only concerns the zero and
one-excitation subspaces. The acceptor is placed at the center of
ring. Since the hoping between the non-adjacent sites would be weak,
we consider only the adjacent hopping of excitation on the ring with
strength $g$. As we know, the visible light takes main part of the
energy of the solar spectrum and the corresponding wavelength is about
$5\times10^{3}\dot{\mathrm{A}}$. In the cell of photosynthesis bacteria,
the B800 and B850 ring usually have the radius about $4-6\mathrm{nm}$
\cite{HuBJ1998}. Thus, the BChl molecules, the unit of B800 and B850
ring, are coupled uniformly to the incident light of frequency $\omega$
which is described by the creation (annihilation) operator $b^{\dagger}$($b$).

To mimic the major function of natural process, we adopt the uniform
couplings in the present model, namely, $\epsilon_{i}=\epsilon$.
Then the model Hamiltonian reads as \begin{equation}
H=H_{D}+\epsilon_{A}A^{\dagger}A+H_{DA}+\omega b^{\dagger}b+J\sum_{i=1}^{N}\left(e_{i}^{+}b+\mathrm{h.c}\right),\end{equation}
where $H_{D}=\sum_{i=1}^{N}\left[\epsilon e_{i}^{\dagger}e_{i}+g\left(e_{i}^{\dagger}e_{i+1}+\mathrm{h.c.}\right)\right]$
and $H_{DA}=\sum_{i=1}^{N}t_{0}\left(e_{i}A^{\dagger}+\mathrm{h.c.}\right)$
with $e_{i}^{\dagger}=\left\vert e\right\rangle _{i}\left\langle g\right\vert $
and $A^{\dagger}=\left\vert e\right\rangle _{A}\left\langle g\right\vert $.
The physical implementation of the present structure could be quantum
dot, which has the size about $2-10\mathrm{nm}$ and distance about
$10\mathrm{nm}$ \cite{QdotsReview}. The recent experimental synthesis
of 12-porphyrin ring \cite{nanoring2011exp} also opens up the possibilities
of designing resemblance of natural light-harvesting element. The
generic model with $N$ donors and $M$ acceptors has been discussed
with the master equation \cite{prb782008}. The similar setup with
all donors and acceptors in a chain has also been discussed to reveal
the optimal constitution of the two components \cite{CaoJPCB2010}.

In this letter, we only consider the effective transferring process
and explore the advantage of the spatial configuration of the ring
type. The collective excitation of the donor ring is described by
the Fourier transformation $e_{j}=\sum_{k}e^{ikj}\tilde{e}_{k}/\sqrt{N}$
and $e_{j}^{\dagger}=\sum_{k}e^{-ikj}\tilde{e}_{k}^{\dagger}/\sqrt{N}$,
where $\tilde{e}_{k}\left(\tilde{e}_{k}^{\dagger}\right)$ is the
annihilation (creation) operator of the collective mode with definite
momentum $k$. Here, the summation is over all the discrete momentum
$k_{n}=2\pi\left(n-1\right)/N$, with $n=1,...,N$. Indeed, in the
large $N-$limit, we can show that $[\tilde{e}_{k},\tilde{e}_{k'}^{\dagger}]\rightarrow\delta_{kk'}$,
thus the collective excitations behave as bosons \cite{sunprl}. In
terms of the boson-like operators $\tilde{e}_{k}\left(\tilde{e}_{k}^{\dagger}\right)$,
the donor Hamiltonian is seemly diagonalized as $H_{D}=\sum_{k}\left(\epsilon+2g\cos k\right)\tilde{e}_{k}^{\dagger}\tilde{e}_{k}$,
which represents an energy band with $N$ sub-energy-levels. We note
that interaction term is rewritten as $H_{DA}=\sqrt{N}t_{0}\left(\tilde{e}_{0}A^{\dagger}+\mathrm{h.c.}\right)$,
which shows that only the zero mode excitation with $k=0$ is coupled
to the acceptor while the others described by $\tilde{e}_{k}\left(\tilde{e}_{k}^{\dagger}\right)$
($k\neq0$) are decoupled with the acceptor. Therefore, the above
Fourier transformation separates the total Hamiltonian into two un-coupled
parts, $H^{\prime}=H_{D}-\left(\epsilon+2g\right)\tilde{e}_{0}^{\dagger}\tilde{e}_{0}$
and \begin{eqnarray}
H_{\mathrm{eff}} & = & \omega_{0}\tilde{e}_{0}^{\dagger}\tilde{e}_{0}+\omega b^{\dagger}b+\omega_{A}A^{\dagger}A\nonumber \\
 &  & +\sqrt{N}[(t_{0}\tilde{e}_{0}^{\dagger}A+J\tilde{e}_{0}^{\dagger}b)+\mathrm{h.c.}],\end{eqnarray}
where $\omega_{0}=\epsilon+2g-i\kappa$ and $\omega_{A}=\epsilon_{A}-i\Gamma$.
We need to point out that the dissipation rates $\kappa$ and $\Gamma$
have been phenomenologically introduced to describe the loss of excitations
from the donors and the acceptor respectively.

Next we give two remarks on the implication of the above effective
Hamiltonian: 1. the zero mode excitation described by $\tilde{e}_{0}^{\dagger}$
has the energy $E_{k=0}=\epsilon+2g$. Acting on the ground state
$\left\vert 0\right\rangle \equiv\left\vert g_{1},g_{2},..,g_{N}\right\rangle $,
$\ \tilde{e}_{0}^{\dagger}$\ gives a uniform superposition $\left\vert 1\right\rangle \equiv\left\vert 1_{k=0}\right\rangle =\sum_{j=1}^{N}\left\vert 1_{j}\right\rangle /\sqrt{N}$
of the single localized excitations $\left\vert 1_{j}\right\rangle \equiv\left\vert g_{1},...,g_{j-1},e_{j},g_{j+1},...,g_{N}\right\rangle $
in the $j$th donor ($j=1,2,...,N$). It is similar to the single
magnon state in the spin wave system. It has been numerically proved
that the collective excited initial state $\left\vert 1_{k=0}\right\rangle $
would result in the maximum efficiency with respect to any other mode
\cite{prb782008}. 2. In single excitation subspace, the photon-assisted
donor-acceptor system could be described as a three-level $\Lambda$
system as illustrated in Fig.\ref{Fig1}(b). The dash-boxed area is
the corresponding single excitation subspace. Interestingly, the incident
light only couples to the zero mode of the donor ring. Since all other
modes are decoupled from the capture process, they do not contribute
to the light-harvesting process, thus the excitation energy is transferred
to the acceptor only through the zero mode channel. Therefore, the
transferring efficiency will be improved dramatically, if the nature
light-harvesting systems were optimized to emerge a zero mode.

In the single excitation case, the evolution of excitation is constrained
in the subspace spanned by $\left\vert 1_{b}\right\rangle \equiv\left\vert 1,0,0\right\rangle ,\left\vert 1_{D}\right\rangle \equiv\left\vert 0,1,0\right\rangle $
and $\left\vert 1_{A}\right\rangle \equiv\left\vert 0,0,1\right\rangle $
for the direct product state $\left\vert a,b,c\right\rangle \equiv\left\vert a\right\rangle \otimes\left\vert b\right\rangle \otimes\left\vert c\right\rangle $
of the photon, donor in zero-mode and acceptor respectively. Let $\left\vert \phi\left(t\right)\right\rangle $
be the single excitation wave function with corresponding amplitudes
$u\left(t\right)$, $v\left(t\right)$ and $w\left(t\right)$ to the
above basis vectors. The Schrodinger equation is reduced into $i\dot{\mathcal{V}}\left(t\right)=\mathbf{M}\mathcal{V}\left(t\right)$
for $\mathcal{V}\left(t\right)=\left[u\left(t\right),v\left(t\right),w\left(t\right)\right]^{\mathbf{T}}$
and \begin{equation}
\mathbf{M}=\left(\begin{array}{ccc}
\omega & \sqrt{N}J & 0\\
\sqrt{N}J & \omega_{0} & \sqrt{N}t_{0}\\
0 & \sqrt{N}t_{0} & \omega_{A}\end{array}\right).\end{equation}

\begin{figure}
 \includegraphics[width=7cm]{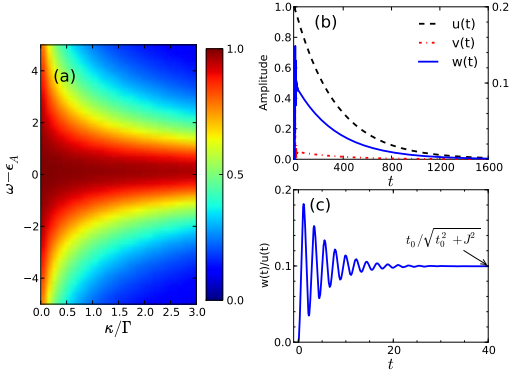} \caption{(\textit{Color online}) (a) Transfer efficiency $\eta$ vs detuning
$\omega-\epsilon_{A}$ and dissipation parameter $\kappa/\Gamma$.
The slow change of peak value indicates that the efficiency at resonance
is insensitive to the donor decay rate $\kappa$. (b) Amplitude evolution
$\left|u\left(t\right)\right|$, $\left|v\left(t\right)\right|$ and
$\left|w\left(t\right)\right|$, (c) Amplitude ratio $\left|w\left(t\right)/u\left(t\right)\right|$
at resonance $\omega=\epsilon_{A}$ and $\kappa=\Gamma/3$.}

\label{Fig:2} 
\end{figure}

The energy transfer is usually understood as the decay from the donors
in excited states \cite{prb782008,Aspuru-Guzik_JCP08_09}, thus the
overall transfer efficiency is given by an integral $\eta=\int_{0}^{\infty}2\Gamma\left\vert v\left(t\right)\right\vert ^{2}dt.$
In practice, with the hopping from the donor to the acceptor, the
rate of transferring excitation to outside agent should be larger
than that only by the dissipation from the excited donors, i.e., $\sqrt{N}t_{0}\gg\kappa$
and $\Gamma>\kappa$. Typically, we will choose the parameters \cite{HuBJ1998}
here as $\epsilon_{A}/t_{0}=10$, $g/t_{0}=0.3$ and $\Gamma/t_{0}=0.3$
with $t_{0}\approx10\mathrm{ps}^{-1}$. In most cases, we set the
total donor number $N=8$. We illustrate the dependence of efficiency
on detuning $\omega-\epsilon_{A}$ and dissipation parameter $\kappa/\Gamma$
in Fig.\ref{Fig:2}(a). There exists a high peak at $\omega=\epsilon_{A}$
for given $\kappa/\Gamma$ with the peak value of efficiency almost
unchanged. This observation reflects the efficiency is insensitive
to the decay of the donor ring. At resonance, the excitation amplitude
of the donor is highly suppressed, as illustrated in Fig.\ref{Fig:2}(b).
For the system comprises only the donor and the acceptor with initial
excitation on the donor, there is a up-bound for the overall transfer
efficiency $\eta_{\mathrm{max}}^{\prime}=\Gamma/\left(\Gamma+\kappa\right)$.
In the present discussion with the light capture included, the efficiency
actually goes beyond the up-bound.

The above discovery that the transfer efficiency is insensitive to
the noisy of the donors at resonance can be explained according to
the dark state, which has been widely investigated in quantum optics
\cite{Scullybook}. To this end, we write the dark state $\left\vert D_{0}\left(t\right)\right\rangle =\cos\theta\left(t\right)\left\vert 1_{b}\right\rangle -\sin\theta\left(t\right)\left\vert 1_{A}\right\rangle $
in the single-excitation subspace for $\Gamma\!\!=\!\!0$, where $\theta\left(t\right)=\arctan\left[J/t_{0}\exp\left[-i\left(\epsilon_{A}-\omega\right)t\right]\right]$.
$\left\vert D_{0}\left(t\right)\right\rangle $ is an eigenstate with
vanishing eigen-value, namely, $H_{\mathrm{eff}}\left\vert D_{0}\left(t\right)\right\rangle =0$.
It has been demonstrated that the perfect transfer of population between
the two low-lying energy levels can be achieved by adiabatically tuning
the Rabi frequencies $J$ and $t_{0}$ \cite{DarkS}. In this process,
the excitation on the upper energy level is suppressed to avoid the
dissipation of excitations.

This dark stated based mechanism persists in the present artificial
system with bionic structure. For this system, the evolution wave
function is

\begin{equation}
\left\vert \phi\left(t\right)\right\rangle =\sum_{i=1}^{3}\frac{e^{-ix_{i}t}\sqrt{\mathcal{N}_{i}}}{\prod_{j\neq i}\left(x_{i}-x_{j}\right)}\left\vert E_{i}\right\rangle ,\end{equation}
where $\left\vert E_{i}\right\rangle =\mathcal{N}_{i}^{-1/2}[\left(\left(x_{i}-\omega_{0}\right)\left(x_{i}-\omega_{A}\right)-Nt_{0}^{2}\right)\left\vert 1_{b}\right\rangle -\sqrt{N}J\left(x_{i}-\omega_{A}\right)\left\vert 1_{D}\right\rangle +NJt_{0}\left\vert 1_{A}\right\rangle ]$
is an eigenstate with a normalized constant $\mathcal{N}_{i}^{2}=N^{2}J^{2}t_{0}^{2}+NJ^{2}\left\vert x_{i}-\omega_{A}\right\vert ^{2}+\left\vert \left(x_{i}-\omega_{0}\right)\left(x_{i}-\omega_{A}\right)-Nt_{0}^{2}\right\vert ^{2}$;
$x_{i}$ is the corresponding eigenvalue of the matrix $\mathbf{M}$,
whose expressions are not explicitly written down since they are too
lengthy. If we choose the parameters as previous $\epsilon_{A}/t_{0}=10$,
$g/t_{0}=0.3$, $\kappa/t_{0}=0.1$, $\Gamma/t_{0}=0.3$ and $J/t_{0}=0.1$,
the eigen states at resonance $\omega=\epsilon_{A}$ are written as
\begin{eqnarray}
\left\vert E_{1}\right\rangle  & \simeq & -0.995\left\vert 1_{b}\right\rangle +0.01i\left\vert 1_{D}\right\rangle +0.01\left\vert 1_{A}\right\rangle ,\nonumber \\
\left\vert E_{2}\right\rangle  & \simeq & 0.07\left\vert 1_{b}\right\rangle +\left(0.7+0.03i\right)\left\vert 1_{D}\right\rangle +0.7\left\vert 1_{A}\right\rangle ,\nonumber \\
\left\vert E_{3}\right\rangle  & \simeq & 0.07\left\vert 1_{b}\right\rangle +\left(-0.7+0.03i\right)\left\vert 1_{D}\right\rangle +0.7\left\vert 1_{A}\right\rangle .\end{eqnarray}
Here, $\left\vert E_{1}\right\rangle $ is the dark state with very
small component in donor excitation, which is proportional to $J\Gamma/\sqrt{N}t_{0}^{2}$.
The initial one photon state can be rewritten with the above eigenstates
as the basis, i.e., $\left\vert 1\right\rangle _{b}=0.995\left\vert E_{1}\right\rangle +0.07\left\vert E_{2}\right\rangle +0.07\left\vert E_{2}\right\rangle $.

The component of dark state in the initial state is approximate $p_{E_{1}}=t_{0}/\sqrt{t_{0}^{2}+J^{2}}$,
which is almost $1$ under the practical condition $J\ll t_{0}$.
In the capture and transferring process, the population on excited
donors is suppressed to be small in avoiding dissipation, as illustrated
in Fig.\ref{Fig:2}(b). On the dark state, the system decays very
slowly at the rate $\gamma_{D}\simeq\Gamma\left(J/t_{0}\right)^{2}$,
while it dissipates quickly on the bright state ($\left\vert E_{2}\right\rangle $,
$\left\vert E_{3}\right\rangle $) at rate $\gamma_{B}\simeq\left(\kappa+\Gamma\right)/2$.
Thus, for large time scale $\tau_{B}>1/\gamma_{B}$, the main contribution
of transferring is carried on by the dark state. We demonstrate the
ratio of populations on the photonic state and on acceptor in Fig.\ref{Fig:2}(c).
The asymptotic value of this ratio for long time is approximately
$t_{0}/\sqrt{t_{0}^{2}+J^{2}}$, which is the one of dark state in
the initial state.

\begin{figure}
 \includegraphics[width=6cm]{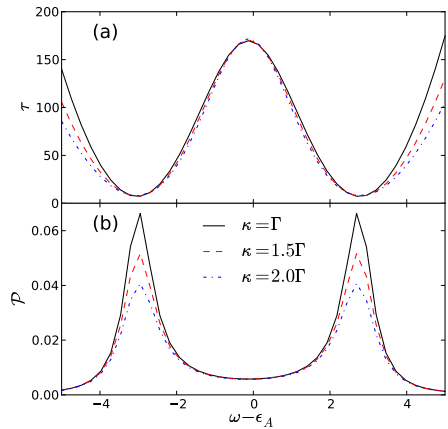} \caption{(\textit{Color online}). (a) Average transfer time vs frequency detuning
$\omega-\epsilon_{A}$. (b) mean power output $\mathcal{P}$ vs detuning
$\omega-\epsilon_{A}$ for different dissipation rate on the donor
ring $\kappa=\Gamma$, $1.5\Gamma$ and $2.0\Gamma$.}

\label{Fig3} 
\end{figure}

In the transferring process, another important quantity is the average
transfer time, which is defined as $\tau=\eta^{-1}\int_{0}^{\infty}2\Gamma t\left\vert v\left(t\right)\right\vert ^{2}dt.$
To effectively utilize the energy, the excitation should be transferred
with a high efficiency and also within a short time scale. In the
previous discussions, we have proved that the efficiency can be improved
via the dark state mechanism. However, we have illustrated that the
decay rate of photon at resonance is suppressed by a factor $\left(J/t_{0}\right)^{2}$,
while the efficiency is proportional to $[1+\left(J/t_{0}\right)^{2}]^{-1/2}$.
We meet a dilemma that the efficiency and the average transfer time
can not be optimized simultaneously. In Fig.\ref{Fig3}(a), we demonstrate
the average transfer time as a function of the dissipation rate of
the donor and the detuning. The small peak in the center hints that
the transfer time is not optimal at resonance $\omega=\epsilon_{A}$.
It is readily seen in Fig.\ref{Fig3}(a) that the optimal frequencies
of quick transfer are not at resonance $\left(\omega=\epsilon+2g\right)$
but split into two, which is known as the Rabi splitting in quantum
optics. The two peaks can be determined by exactly diagonalizing the
photon assisted donor-acceptor system in single excitation subspace
as $\omega_{\pm}=\left(\epsilon_{A}+2g+\epsilon\right)/2\pm\lbrack\left(\epsilon+2g-\epsilon_{A}\right)^{2}/4+Nt_{0}^{2}]^{1/2}$.
The transfer time reaches its minimum optimal point when $\omega=\omega_{\pm}$,
while the efficiency is not at its maximum as illustrated in Fig.\ref{Fig3}.

In fact, such dilemma has been met in many investigations about heat
engines \cite{Carnot1975}: the Carnot heat engine converts the heat
into work with maximum efficiency, while it takes infinite long time.
In practice, one would concern more about the output power, which
characterizes the output energy within unit time. In the present case,
we introduce a similar quantity $\mathcal{P}=\eta/\tau$ characterizing
the ability of the excitation energy transfer, which is called the
mean transfer power. In Fig.\ref{Fig3} (b), we demonstrate the mean
power $\mathcal{P}$ as a function of dentuning between the incident
light frequency and the acceptor excitation energy for different donor
dissipation rates $\kappa=\Gamma$, $1.5\Gamma$ and $2\Gamma$. The
optimal frequency is $\omega=9.1t_{0}$ and $\omega=14.7t_{0}$, which
are different from the optimal value of both efficiency and average
transfer time. Actually, we have observed that the photosynthesis
systems choose to absorb sunlight mainly from two domains of the solar
spectrum. For example, the green plants mainly use red and blue color
photons. The present model can be utilized to justify this observation.
However, for the practical system, the estimation of the exactly frequency
goes beyond the scope of the present model because it concerns very
complicated biological environments.

In summary, we have studied the light capture and excitation transfer
process in a generic model consisted of donor and acceptor assisted
by photons. By optimizing this artificial photosynthesis system to
realize an effective three-level $\Lambda$ configuration evolving
dark state, we demonstrated the coherent population transferring through
the dark state channels, where the dissipation from donors is effectively
suppressed. In the present studies, we deal with the dissipation of
the excitation by phenomenologically introducing an imaginary part
to the Hamiltonian of the donor. However, in reality, this dissipation
is always connected to the vibration degrees of freedom, which may
be account for the dimerized structure as discussed in the LHC I and
LHC II \cite{SYangJCP2010}. And also, it is worth to investigate
effect of vibrations in designing some artificial light-harvesting
systems. For the present model, we can also discuss the quantum or
classical correlations of the output excitations with more than one
photon, which could be used to explain the mechanism of the light-harvesting
system avoiding the damage from high intensity light source.

The work is supported by National Natural Science Foundation of China
a under Grant Nos. 10935010 and 11074261.


\begin{thebibliography}{24}
\bibitem{energypro} G.R. Fleming and M.A. Ratner, Phys. Today 61,
28(2008); A.C Benniston and A. Harriman, Mater. Today 11, 26 (2008).

\bibitem{qcoherenceexp} G.S. Engel, et al., Nature 446, 782 (2007);
H. Lee, et al., Science 316, 1462(2007); E. Collini, et al. , Nature
463, 644 (2010). T. R. Calhoun et al., J. Phys. Chem. B 113, 16291(2009).
G. Panitchayangkoon et al., arXiv:1001.5108.

\bibitem{YZhao} Y. Zhao et al, J. Phys. Chem. B \textbf{103}, 3854(1999);
Y. Zhao et al, Phys. Rev. E \textbf{69}, 032902 (2004).

\bibitem{prb782008} A. Olaya-Castro, C. F. Lee, F. F. Olsen and N.
F. Johnson, Phys. Rev. B \textbf{78}, 085115(2008).

\bibitem{venturi08cpc} V. Balzani, A. Credi and M. Venturi, Chem.
Phys. Chem. \textbf{1}, 26(2008).

\bibitem{Guzik09JPCB} P. Rebentrost, M. Mohseni, and A. Aspuru-Guzik,
J. Phys. Chem. B \textbf{113}, 9942 (2009).

\bibitem{mukamel09JCP} B. Palmieri, D. Abramavicius, and S. Mukamel,
J. Chem. Phys. \textbf{130}, 204512 (2009).

\bibitem{Thorwart09CPL} M. Thorwart, J. Eckel, J.H. Reina, P. Nalbach
and S. Weiss, Chem. Phys. Lett. \textbf{478}, 234 (2009).

\bibitem{plenio09JCP} F. Caruso, A. W. Chin, A. Datta, S. F. Huelga,
and M. B. Plenio, J. Chem. Phys. \textbf{131}, 105106 (2009).

\bibitem{Guzik2010APL} A. Perdomo, L. Vogt, A. Najmaie and A. Aspuru-Guzik,
App. Phys. Lett. \textbf{96}, 093114(2010).

\bibitem{Sarovar10NP} M. Sarovar, A. Ishizaki, G. R. Fleming, and
K. B. Whaley, Nat. Phys. \textbf{6}, 462 (2010);

\bibitem{Ishizaki10NJP} A. Ishizaki and G.R. Fleming, New. J. Phys.
\textbf{12}, 055004(2010).

\bibitem{CaoJPCB2010} J-H. Kim and J-s. Cao, J. Phys. Chem. B, 114,
16189 (2010).

\bibitem{Aspuru-Guzik_JCP08_09} M. Mohseni, P. Rebentrost, S. Lloyd,
and A. Aspuru-Guzik, J. Chem. Phys. \textbf{129}, 174106 (2008); P.
Rebentrost, M. Mohseni, I. Kassal, S. Lloyd, and A. Aspuru-Guzik,
New J. Phys. \textbf{11}, 033003 (2009).

\bibitem{SYangJCP2010} S. Yang, D.Z. Xu and C.P. Sun, J. Chem. Phys.
\textbf{132}, 234501 (2010).

\bibitem{x-rays} G. McDemott et al, Nature \textbf{374}, 517(1995);
J. Koepke et al, Structure \textbf{4}, 581(1996).

\bibitem{HuBJ1998} X. Hu and K. Schulten, Biophys. J. \textbf{75},
683 (1998).

\bibitem{Hu97PT} X. Hu and K. Schulten, Phys. Today. \textbf{50},
28 (1997).

\bibitem{QdotsReview} S.M. Reimann and M. Manninen, Rev. Mod. Phys.
\textbf{74}, 1283 (2002).

\bibitem{nanoring2011exp} M. C. O'Sullivan, et al, Nature \textbf{469},
72(2010).

\bibitem{sunprl} C. P. Sun, Y. Li, and X.F. Liu, Phys. Rev. Lett
\textbf{91}, 147903 (2003)

\bibitem{Scullybook} M. O Scully, \textit{Quantum Optics}, Cambridge
University Press(1997)

\bibitem{DarkS} F.T. Hioe, Phys. Lett. A \textbf{99}, 150(1983);
F.T. Hioe and J.H. Eberly, Phys. Rev. A \textbf{29}, 690(1984); J.R.
Kuklinski, et.al, Phys. Rev. A \textbf{40}, 6741(1989).

\bibitem{Carnot1975} F.L. Curzon and B. Ahlborn, Am. J. Phys. \textbf{43},
22(1975). 
\end{thebibliography}
\end{document}